\documentclass[preprint2]{aastex}
\usepackage{lscape}
\usepackage{epsfig}

\def\kms{\relax \ifmmode {\,\rm km\,s}^{-1}\else \,km\,s$^{-1}$\fi}

\def\farcs{\hbox{$.\!\!^{\ prime\prime}$}}

\def\arcmin{\hbox{$^\prime$}}
\def\arcsec{\hbox{$^{\prime\prime}$}}
\def\secd#1.#2{ #1\farcs#2 } 

\def\mincir{\ \raise-2.truept\hbox{\rlap{\hbox{$\sim$}}\raise5.truept
    \hbox{$<$}\ }}
\def\magcir{\ \raise-2.truept\hbox{\rlap{\hbox{$\sim$}}\raise5.truept
    \hbox{$>$}\ }}

\def\nii{[N {\sc ii}]}
\def\sii{[S {\sc ii}]}

\def\oii{[O {\sc ii}]}
\def\oi{[O {\sc i}]}

\def\oiii{[O {\sc iii}]}
\def\neiii{[Ne {\sc iii}]}
\def\cliii{[Cl {\sc iii}]}
\def\ha{H$\alpha$}
\def\hb{H$\beta$}
\def\ji{$\rm{J}_1$}
\def\jii{$\rm{J}_2$}
\def\ki{$\rm{K}_1$}
\def\kii{$\rm{K}_2$}
\def\kiii{$\rm{K}_3$}
\def\kiv{$\rm{K}_4$}
\def\ri{$\rm{R}_1$}
\def\rii{$\rm{R}_2$}
\def\chb{$c_{\rm H\beta}$}

\shorttitle{Jets and knots of NGC~7009}
\shortauthors{Gon\c calves et al.}

\begin{document}

\title{The physical parameters, excitation 
       and chemistry of the rim, jets and knots of the planetary nebula NGC~7009 
      \footnote{Based on observations obtained at the 2.5 m Isaac Newton 
Telescope (INT) of the European Northern Observatory, and with the NASA/ESA
{\it Hubble Space Telescope}, obtained at the Space Telescope Science
Institute, which is operated by AURA for NASA under contract NAS5-26555.}}

\author{Denise R. Gon\c calves}

\affil{Instituto de Astrof\'{\i}sica de Canarias,
      \\E-38205 La Laguna, Tenerife, Spain
      \\e-mail: denise@ll.iac.es}

\author{Romano L. M. Corradi}

\affil{Isaac Newton Group of Telescopes,
       \\ Apartado de Correos 321, E-38700, Sta. Cruz de la Palma, Spain
       \\e-mail: rcorradi@ing.iac.es}

\author{Antonio Mampaso}

\affil{Instituto de Astrof\'{\i}sica de Canarias,
       \\ E-38205 La Laguna, Tenerife, Spain
       \\e-mail: amr@ll.iac.es}

\author{Mario Perinotto}

\affil{Dipartimento di Astronomia e Scienza dello Spazio, Universit\`a di Firenze, 
       \\Largo E. Fermi 5, 50125 Firenze, Italy 
       \\e-mail: mariop@arcetri.astro.it}

\begin{abstract}

We present long-slit optical spectra along the major axis of the planetary
nebula NGC~7009.  These data allow us to discuss the physical, excitation and
chemical properties of all the morphological components of the nebula, including
its remarkable systems of knots and jets.  The main results of this analysis are
the following:  {\it i)} the electron temperature throughout the nebula is
remarkably constant, $T_e$\oiii\ = $10\,200$~K; {\it ii)} the bright inner rim
and inner pair of knots have similar densities of $N_e \sim 6000$~cm$^{-3}$,
whereas a much lower density of $N_e \sim$ 1500~cm$^{-3}$ is derived for the
outer knots as well as for the jets; {\it iii)} all the regions (rim, inner
knots, jets and outer knots) are mainly radiatively excited; and {\it iv)}
there are no clear abundance changes across the nebula for He, O, Ne, or S.
There is a marginal evidence for an overabundance of nitrogen in the outer knots
(ansae), but the inner ones (caps) and the rim have similar N/H values that are
at variance with previous results.  Our data are compared to the predictions of
theoretical models, from which we conclude that the knots at the head of the
jets are not matter accumulated during the jet expansion through the
circumstellar medium, neither can their origin be explained by the proposed HD
or MHD interacting-wind models for the formation of jets/ansae, since the
densities as well as the main excitation mechanisms of the knots, disagree with
model predictions.

\end{abstract}


\section{Introduction}

In addition to their large-scale structures, mainly identified in the 
forbidden [OIII] emission line, many planetary nebulae (PNe) have
a number of smaller-scale structures which are
instead more prominent in low-ionization lines, such as [N~{\sc ii}],
\oii\ and \sii. In a previous paper (Gon\c calves, Corradi, \& Mampaso
2001), we have used the acronymous LISs to identify these
low-ionization structures.  LISs appear with different morphological
and kinematical properties in the form of pairs of knots, filaments,
jets, or isolated features moving with supersonic velocities through
the large-scale components in which they are located, or instead as
structures with the above morphologies but with low velocities that
do not differ substantially from that of the main nebula\footnote{The 
fast, low-ionization emission regions, FLIERs (Balick et al. 1993) and 
bipolar, rotating, episodic jets, BRETs (L\'opez, V\'azquez, \& 
Rodr\'\i guez 1995) are particular types of LISs, i.e., high-velocity 
knots and jets.}. LISs can be easily seen in the
imaging catalogues of Balick (1987), Schwarz, Corradi, \& Melnick
(1992), Manchado et al.  (1996), Corradi et al. (1996) and G\'orny et
al. (1999). For a detailed study concerning the morphological and
kinematical classification of LISs in PNe, see Gon\c calves et al.~(2001).

A very interesting class of LISs are jets. These are highly collimated
outflows (with aspect ratios varing between 3 and 20; L\'opez 1997), 
which are directed  radially outward from the central
star, appear in opposite symmetrical pairs, and move with velocities
substantially larger than those of the ambient gas that forms the main
bodies of the nebulae. It should be noted in addition that jets are 
much more collimated than other bipolar outflows of PNe, i.e, more 
collimated than the lobes of a bipolar PN.

NGC~7009, the ``Saturn Nebula'', is a well studied elliptical PN, which
possesses a jet-like system as well as two pairs of low-ionization
knots along its major axis.  A number of previous studies showed that
high-excitation lines  dominate the inner regions along the minor
axis, while the low-excitation ones are enhanced at the ends of the
major axis. The ionization structure is further enriched by the
fact that the low-excitation region present strong variations in 
excitation level and clumpiness. NGC~7009 was classified as an oxygen-rich 
PN (Hyung
\& Aller 1995), with an O/C ratio exceeding 1, and anomalous N, O, and C
abundances (Baker 1983; Balick et al.~1994, Hyung \& Aller 1995).  Its
central star is an H-rich O-type star, with effective temperature of
82\,000~K (M\'endez, Kudritzki, \& Herrero 1992; Kingsburgh \& Barlow
1992).  The kinematics of NGC~7009 was studied by Reay \& Atherton
(1985) and Balick, Preston \& Icke (1987).  The distance
determinations of NGC~7009 vary from 0.5 to 2.3 kpc from various
indirect methods (Acker et al. 1992; Cahn, Kaler \& Stanghellini 1992;
Maciel 1995).

According to the theoretical models for the formation of jets and of
the pairs of knots that are often found at their tips (Soker 1990;
Garc\'\i a-Segura 1997; Garc\'\i a-Segura et al.~1999; Garc\'\i
a-Segura \& L\'opez 2000; Steffen, L\'opez, \& Lim 2001; Gardiner \&
Frank 2001; Blackman, Frank, \& Welch 2001), a significant density
contrast between the jet itself and the knots at its tip is expected.
This is a quantity that can in principle be determined by optical
spectroscopy.

In addition, previous observational work suggests that the LISs of
NGC~7009 and other PNe have higher abundances of nitrogen than their
main bodies (see Balick et al.~1994 and references therein).  These 
authors proposed that such an overabundance might indicate that the 
knots of NGC~7009 were ejected at a different epoch than the main body 
of the nebula, when the stellar surface was enriched of these elements 
due to the drastic changes of nitrogen in AGB stars and some dredge-up
episode, or that the knots come from chemically peculiar regions in
the stellar surface or its interior.

In this paper, we test the above two issues  by obtaining spectroscopic
data for NGC~7009, deriving its physical and chemical properties,
and finally showing that neither a density contrast between the jets
and knots is found, nor are their abundances  strongly enhanced compared to the
rest of the nebula. In the following sections, we first describe the
data, their reduction and the various morphological/kinematical
structures of the nebulae (Sections 2 and 3). Then we proceed with the
analysis of the PN spectra in terms of physical, excitation and
chemical parameters (Section 4) and, finally, we discuss the
implication of the present results for the formation of the LISs of
NGC~7009 and give our conclusions, in Sections 5 and 6.

\section{Observations and Data Reduction}

Spectra of NGC~7009 (PN G037.7-34.5) were obtained on 2001 August 29
at the 2.5 m Isaac Newton Telescope (INT) at the Observatorio del 
Roque de los Muchachos (European Northern
Observatory, La Palma, Spain), using the Intermediate Dispersion
Spectrograph (IDS).  The IDS was used with the 235~mm camera and the
R300V grating, providing a spectral coverage from 3650 \AA\ to 7000 \AA\
with a spectral reciprocal dispersion of 3.3 \AA~pix$^{-1}$. The
spatial scale of the instrument was 0$''$.70~pix$^{-1}$ with the
TEK5 CCD. Seeing varied from 0$''$.9 to 1$''$.1. The slit width
and length were 1.5\arcsec\ and 4\arcmin, respectively.  The slit was
positioned through the center of the nebula at P.A. = 79$^{\circ}$, and
the exposure times were 3$\times $30~s, 3 $\times $ 120~s, and
3 $\times$ 600~s.

During the night, bias frames, twilight and tungsten flat-field
exposures, wavelength calibrations and exposures of standard stars
(BD+332642, Cyg OB2\#9, HD19445, and BD+254655) were obtained. Spectra
were reduced following the IRAF instructions for long-slit spectra,
being bias-subtracted, flat-fielded, combined in order to improve the
signal-to-noise ratio and eliminate cosmic rays, wavelength
calibrated, and sky-subtracted. Finally, they were flux-calibrated
using the above-mentioned standard stars  and the mean atmospheric extinction
curve for La Palma.

We also retrieved {\itshape HST} \nii\ and \oiii\ archive images of
NGC~7009 obtained on 1996 April 28, and on 2000 April 7, respectively,
with the WFPC2 camera in the F658N and F502N filters, with total
exposure times of 1200~s and 320~s.

\section{The morphological and kinematical components of NGC~7009: 
previous works}

\begin{figure}[ht]
\plotone{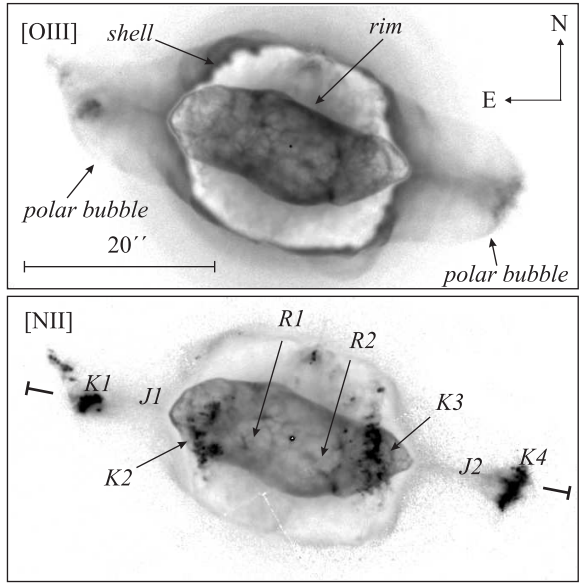}
\caption[ ]{The {\itshape HST} \oiii\ and \nii\ images of NGC~7009 on a logarithmic 
intensity scale.  Slit position is indicated by short lines 
(P.A.  = 79$^\circ$), while labels mark the position of the several 
structures, where ``K'', ``J'' and ``R'' stand for ``knot'',
``jetlike'', and ``rim'', respectively. Extensions of the selected structures with 
respect to the central star are: from $-$26.6$''$ to $-$21.7$''$, \ki; 
from $-$20.3$''$ to $-$16.1$''$, \ji;  from $-$14.7$''$ to $-$6.3$''$, \kii; 
from $-$4.9$''$ to $-$1.4$''$, \ri; from 2.1$''$ to 3.5$''$; \rii; 
from 4.9$''$ to 14.0$''$, \kiii; from 17.5$''$ to 22.4$''$, \jii; 
and from 23.8$''$ to 28.7$''$, \kiv.
\label{I-n7009}}
\end{figure}

\placefigure{I-n7009}

The {\itshape HST} \oiii\ and \nii\ images of NGC~7009 are shown in
Figure~\ref{I-n7009}.  The nebula consists of several components. The
brightest one is the elliptical inner shell ({\it rim}) roughly
elongated in the EW direction, where it extends out to $\pm 13''$ from
the center.  The inner portions of the rim cut by our spectroscopic
slit are indicated in Figure~\ref{I-n7009} (bottom) by the labels
R$_1$ and R$_2$.

Surrounding the rim, a fainter elliptical outer shell ({\it shell}) that shares
its main axis with the rim is also visible in the {\itshape HST} images. In 
\oiii, at fainter intensity levels the shell shows two ``polar''
bubbles. Inside these bubbles, collimated LISs are clearly seen in the
\nii\ images. These are composed of a ``pencil'' of emission (the {\it
jetlike} structures, labeled J$_1$ and J$_2$) that connects the inner
rim with two {\it outer knots} (K$_1$ and K$_4$).  The \nii\ image
also shows another pair of microstructures, which appear as
``compact groups of bright knots'' that we call the {\it inner knots}
(K$_2$ and K$_3$). The extension of all features is given in the caption 
of Figure~\ref{I-n7009}. 

The rim of NGC~7009 is likely to be formed by the interaction between
the fast post-AGB and the slow AGB winds (Kwok, Purton, \& Fitzgerald
1978; Frank \& Mellema 1994; Mellema \& Frank 1995; Zhang \& Kwok
1998, Sch\"onberner 2002).  The cavity delimited by the rim is
filled with a hot ($T=1.8\times 10^6$~K) gas emitting thermal X-rays
and its size is in agreement with that of the optical rim (Guerrero,
Gruendl, \& Chu, 2002) providing a direct proof for the shocked fast
stellar wind in the interacting winds scenario.  The formation of the
shell has been ascribed to the action of the photoionization
front on the AGB matter not yet reached by the shock produced by the
fast wind (e.g., Sch\"onberner 2002).  

The origin and nature of the LISs in NGC 7009 are  rather
unclear at present. The inner and outer pairs of knots have been studied by
many authors (Reay \& Atherton 1985; Balick et al. 1994; Hyung \&
Aller 1995; Lame \& Pogge 1996; Balick et al. 1998), who often called
them {\it caps} and {\it ansae}, respectively.  Reay \& Atherton
(1985) have studied the kinematics of the inner and outer knots in the
\oi\ 6300~\AA\ line using Fabry--Perot data. They derived an expansion
velocity for the inner knots of $\pm38$\kms\ with an inclination with
respect to the line of sight $i\cong 51^{\circ}$, and a velocity of
$\pm60$~\kms\ with $i\cong 84^{\circ}$ for the outer pair.  Note that
the determination of the deprojected velocities and inclination angles
by Reay \& Atherton (1985) is mainly based on the proper motion
measurements of Liller (1965), as the velocity in the plane of the sky
appears one order of magnitude larger than the Doppler shift. We have
estimated the separation of the outer knots in the \nii\ {\itshape HST} image
taken in 1996, finding $\sim$50.8 $\pm$ 0.5 arcsec, the uncertainty
depending on the highly irregular morphology of these knots that makes
it difficult to define a centroid.  This is clearly not consistent
with the extrapolation of Liller's measurements (separation of 48
arcsec in 1965 and apparent expansion of 1.6 arcsec per century for
each knot), making the velocity and inclination figures of Reay \&
Atherton (1985) quite uncertain. Their analysis also depends on the
distance adopted, which is poorly known, as mentioned in the
introduction.

No velocity information is available from the literature for the
jetlike structures connecting the rim to the outer knots. If we assume
a continuity between the velocity of the outer knots and these jetlike
structures, then the latter might be real supersonic jets.  A more
detailed study of the kinematics of NGC~7009, and in particular of the
faint ``pencil'' and large-scale structures surrounding it (shell and
polar bubbles), is clearly needed to understand  its nature better.

\section{Data Analysis}

Line intensities were mainly measured from the deep spectrum with
3$\times$600~s exposure time.  The shorter-exposure spectra were
used to measure the bright \hb, \oiii, \ha, and \nii\ lines that are
saturated in the deep spectra. Measurements are given in Table~1,
which lists the wavelength of the line identification and the
corresponding ion (column 1), the line fluxes measured for the
different spatial regions selected (columns 2 to 9, as defined in 
Fig. 1 and Section 3), and the flux integrated
along the slit for the whole nebula, labeled as NEB (column
10). All fluxes are normalized to $I$(\hb) = 100 in each of the regions
considered. 

Errors in the fluxes were
calculated taking into account the statistical error in the
measurement of the fluxes, as well as systematic errors of the flux
calibrations, background determination, and sky subtraction.
Table~2 gives the estimated accuracy of the measured fluxes for a range of 
line fluxes (relative to \hb) in each of the selected regions. 
Absolute \hb\ fluxes integrated along the slit in each region  
are also given in Table~2.

\subsection{Extinction}

Fluxes were extinction-corrected by using \chb=0.16$\pm$ 0.01 (the
logarithmic ratio between observed and dereddened \hb\ fluxes),
determined from the \ha/\hb\ ratio in the brightest regions of the
nebula (\ri\ and \rii).  For the derivation of \chb, we assumed
$T_{\rm e}$ = 10$^4$~K and $n_{\rm e}$ = 10$^4$~cm$^{-3}$, appropriate for the
values that we have computed in the rim (Section 4.2). Theoretical Balmer 
line ratios 
from Osterbrock (1989) and the reddening law of Cardelli, Clayton, \& 
Mathis (1989) were used.  Our derived value is in fair agreement with the
average value (0.15) reported by other authors: \chb\ = 0.08 $\pm$ 0.03
(Hyung \& Aller 1995); \chb\ = 0.24 $\pm$ 0.04 (Lame \& Pogge 1996);
\chb\ = 0.20 (Liu et al. 1995); \chb\ = 0.10 (Rubin et al. 2002).

Note that we applied \chb\ = 0.16 for the fluxes of all the different
zones under analysis, because of the highest S/N ratio of the rim,
thus avoiding the larger errors in the measured Balmer line ratios of
the fainter zones. This assumption of a constant extinction throughout
the nebula is supported by the work of Bohigas, L\'opez, \& Aguilar
(1994), who reported that the \chb\ for the different structures of
NGC~7009 are always lower than 0.28, implying small corrections to the
observed fluxes.

\subsection{Densities and Temperatures}

\begin{figure}[ht]
\plotone{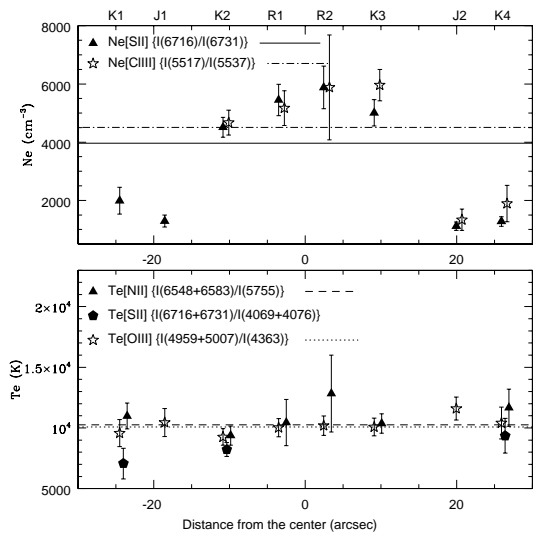}
\caption[ ]{
Electron densities and temperatures determined with different line
ratios as a function of the distance from the center of NGC~7009. The
positions of the various features are marked in the uppermost part of
the plot (see Figure~1). Symbols are plotted slightly displaced in
distance in order to avoid overlapping. The horizontal lines in both
plots give the value of each density/temperature estimate for the
entire nebula: $N_e$\sii(NEB) = 4000 $\pm$ 300~cm$^{-3}$ 
$N_e$\cliii(NEB) = 4500 $\pm$ 400~cm$^{-3}$; $T_e$\oiii(NEB) = 10\,100 $\pm$7 00~K, and 
$T_e$\nii(NEB) = 10\,300 $\pm$ 800~K.
\label{P-temden}}
\end{figure}

As described above, eight different regions were selected in the spectrum of
NGC~7009 (see also Figure~\ref{I-n7009}). The extinction-corrected fluxes
were then used in order to obtain the densities and temperatures for each of
the regions that are plotted in Figure~\ref{P-temden}.  These physical
parameters were obtained using the {\it nebular} package of IRAF (Shaw \&
Dufour 1995), based on the five-level atom program (De Robertis, Dufour, \& Hunt 1987) to
derive the physical conditions in a low-density nebular gas, including
diagnostics for a large number of ions and emission lines.

To estimate densities, we assumed an electron temperature of 10$^4$~K (see
below). In most positions, we were able to measure the densities from both the
\sii\ 6716 \AA/6731 \AA\ and \cliii\ 5517 \AA/5537 \AA\ line ratios.  These  
provide a simultaneous estimate of densities in low-ionization
(\sii) and higher ionization (\cliii) regions. In general, we found a good
agreement between the results from these two density estimators for most of
the structures (see Table~1 and Figure~\ref{P-temden}). Using the total flux 
integrated along the
slit, we obtain $N_e$\cliii/$N_e$\sii\ $\cong$ 1.13; the horizontal
dashed-dashed line in Figure~\ref{P-temden} indicates the $N_e$\cliii\
overall value, while the continuous line shows the corresponding $N_e$\sii\ 
overall density.
 
Figure~\ref{P-temden} clearly shows that the densities of the
outermost knots are very similar to those of the jets connecting them
to the edge of the inner rim, with $N_e$(\ki) = 1900 $\pm$ 500~cm$^{-3}$
and $N_e$(\ji)= 1300$\pm$200~cm$^{-3}$ and, on the opposite side,
$N_e$(\kiv) = 1300 $\pm$ 200~cm$^{-3}$ and
$N_e$(\jii) = 1300 $\pm$ 500~cm$^{-3}$. On the other hand, the inner pair
of knots have higher electron densities, namely
$N_e$(\kii) = 4500 $\pm$ 400~cm$^{-3}$ and 
$N_e$(\kiii) = 5000 $\pm$ 500~cm$^{-3}$, and similar to those of the rim
($N_e$(\ri)= 5500 $\pm$ 600~cm$^{-3}$ and 
$N_e$(\rii) = 5900 $\pm$ 800~cm$^{-3}$).

As for the densities, the temperature estimators used are
appropriate for zones of low- and high-excitation. We used the
following line ratios: $I$(4959 {\AA} + 5007 {\AA})/$I$(4363 {\AA}) for $T_e$\oiii;
$I$(6548 {\AA} + 6583 {\AA})/$I$(5755 {\AA}) for $T_e$\nii ; and 
$I$(6716 {\AA}/6731 {\AA})/$I$(4069 {\AA} + 4076 {\AA}) for $T_e$\sii .  
The general trend of
temperatures (Figure~\ref{P-temden}, bottom) is that, within the errors,
they are constant throughout the  nebula, having an average value
of $T_e$\oiii\ = 10\,200~K and $T_e$\nii\ = 11\,100~K. In spite of a  
small discrepancy of  $T_e$\sii\ at the position of \rii, the temperature 
is remarkably constant across this nebula. 
 
There are a number of papers that discuss the physical properties of
NGC~7009 (Rubin et al. 2002; Hyung \& Aller 1995; Bohigas et al.~1994; 
Balick et al.~1994). In general, there is good
overall agreement between our temperature determinations and previous
ones. 
A more detailed zone-by-zone comparison can
be made with some of these papers.  Balick et al. (1994) studied the 
west rim, cap, and
ansae. They found $T_e$\nii\ and $T_e$\oiii\ of 10\,000~K and 9400~K,
respectively, for the rim, 9600~K and 9600~K for the cap, and 8100~K
and 11\,500~K for the ansa.  With the exception of $T_e$\nii\ of 
that ansa, these figures are very similar to
ours (see Table~1).  From the six regions of the PN studied by
Bohigas et al.~(1994), two coincide with ours (those of
the W and E ansae), and they found, for both structures, an \oiii\
temperature of 9800~K, again consistent, within the errors, with our
measurements. The $T_e$\oiii\ determined from the {\itshape HST}/WFPC2 data  
by Rubin et al.~(2002) is also consistent with the values we 
derived, since they found 9000~K $\leq$ $T_e$\oiii\ $\leq$ 11\,000~K. Their 
STIS long-slit results for temperature are limited to the 13 central 
arcseconds along the major axis. The average $T_e$\oiii\ for  \kii, 
\ri, \rii, and \kiii\ are $\sim$10\,300~K  $\sim$10\,700~K, $\sim$10\,600~K,
 and $\sim$10\,100~K, respectively, which are very similar to our figures 
in Table~1.

The  above papers also report density estimates, whose agreement with our
measurements is  not, however, as good as for the electron temperatures.  For
instance, Balick et al. (1994) found Ne\sii\ = 4900~cm$^{-3}$,
4100~cm$^{-3}$, and 1000~cm$^{-3}$ for the western rim, cap, and ansae,
respectively. They also found a large difference between the
$N_e$\sii/$N_e$\cliii\ ratio at the position of the W rim and cap, which was
measured to be $\sim$3 and $\sim$0.4, respectively, while ours is
$\sim$1.0 and $\sim$0.8 (in line with $N_e$\sii\ $\approx$ $N_e$\cliii\
found by Copetti \& Writzl 2002 for a large sample of PNe). Lame \&
Pogge (1996) found that the rim plus inner knots of NGC~7009 have
densities as high as 8000~cm$^{-3}$, and that the rim without including
\kii\ and \kiii\ has a significantly lower density,
$\sim$4000~cm$^{-3}$. There is also only  moderate agreement between
our densities and those of Bohigas et al.~(1994), who
obtained 2300 $\pm$ 600~cm$^{-3}$ and 4300 $\pm$ 2700~cm$^{-3}$ for
 the W and E ansae, respectively.  The reason of the discrepancies
between different authors might be  the S/N ratio of the relevant
lines, and the seeing.  Note that our observations are, in general, 
much deeper and were obtained during better seeing conditions than the 
above studies, allowing a more precise definition of the spatial regions.

\subsection{Excitation Properties}

\begin{figure}[ht]
\plotone{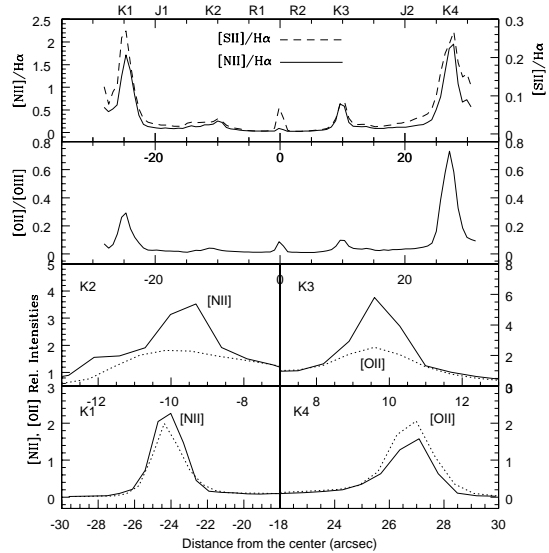}
\caption[ ]{Emission line spatial distribution along P.A. = 79$^\circ$.
From top to bottom: the \nii/\ha\ (continuous) and \sii/\ha\
(dashed) line ratios; the \oii/\oiii\ line ratio; the \nii\ and \oii\
line profiles for the inner (K$_2$, K$_3$) and outer (K$_1$, K$_4$)
pairs of knots.  The position of all the selected features of Figure~1
are marked at the upper part of the plot.
\label{P-ratio}}
\end{figure}

Concerning the general excitation structure of NGC~7009, the
low-ionization (\oi, \nii, \sii, \oii) line emission has local maxima
at the position of the pairs of knots (K$_{1-4}$). The higher (\oiii,
\neiii) ionization emission lines, however, are much more prominent
within the rim of the nebula.

As the line ratio profiles might provide additional interesting
information about the excitation conditions of a
nebula, we show in Figure~\ref{P-ratio} the \nii/\ha, \sii/\ha,
\oii/\oiii, and \nii, \oii\ spatial profiles along the major axis of
the nebula. The top panel clearly shows  the inner and outer pairs of
knots as prominent zones of low ionization. Moreover, the fact that
\nii/\ha, \sii/\ha\ and \oii/\oiii\ ratios are very similar in shape
suggests that their local maxima might not be due to the overabundance
of a given element. These line ratios are rather low at the position
of the jets and the rim implying that the
excitation of the jets is quite different from that of the knots. 
The average values of \nii/\ha\ in the outer knots are
1.00 (\ki) and 0.96 (\kiv), much higher than those usually found for
spherical and elliptical PNe (e.g., Aller \& Liller 1968).  The inner
pair of knots, on the other hand, have less extreme \nii/\ha\ line
ratios, namely 0.13 and 0.14 for \kii\ and \kiii, respectively.

The lower two panels of Figure~\ref{P-ratio} show the \nii\ and \oii\
spatial profiles at the position of the inner (top) and outer (bottom)
pairs of knots.  The \nii\ and \oii\ profiles have been normalized in
the nebular regions just ahead of the knots in order to see whether
the variation of the \nii\ and \oii\ emission through the knots shows
any effects related to collisional quenching of the emission lines.  The 
critical density for collisional de-excitation is $16\,000$~cm$^{-3}$ and 
$3\,100$~cm$^{-3}$ for the \oii\ 3726~\AA\, 3729~\AA\ doublet, respectively, 
while it is $86\,000$~cm$^{-3}$ for the
\nii\ 6583~\AA\ line. Figure 3 shows that the \nii\ and \oii\
profiles are very similar in the outer, lower density (N$_e$ $\le$ 2000~cm$^{-3}$) 
knots, whereas in both inner knots (N$_e$ $\approx$ 5000~cm$^{-3}$) on the 
contrary, the \oii\ profile is much flatter than the \nii\ one. 
This represents a direct proof that collisional quenching is indeed affecting 
the \oii\ doublet in regions of relatively high densities, and caution should be 
exercised when, for example, computing densities using the \oii\ doublet.

\begin{figure}[ht]
\plotone{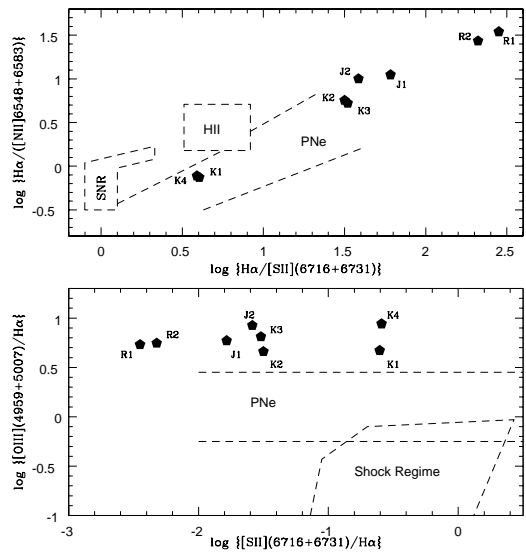}
\caption[ ]{Diagnostic diagrams showing the loci of the different observed 
structures in NGC~7009: {\it Top}, $\log$ \ha/\nii(6548 + 6583) versus $\log$ 
\ha/\sii(6716 + 6731), from Cant\'o (1981). 
{\it Bottom}, $\log$ \oiii(4959 + 5007)/\ha\ versus 
$\log$ \sii(6716+6731)/\ha\ from Phillips \& Cuesta (1999). 
\label{P-shockphot}}
\end{figure}

Peculiar line ratios from structures like the knots of
NGC~7009 have been attributed to anomalous abundances and/or shock
excitation (Balick et al.~1987; Miranda \& Solf 1992). In order to
test whether shock excitation plays a role in NGC~7009,
we show in Figure~\ref{P-shockphot} two relevant diagnostic diagrams
(c.f. Cant\'o 1981; Phillips \& Cuesta 1999).  The behavior of the
different components of NGC~7009 in the \ha/\nii\ (6548 + 6583) versus
\ha/\sii\ (6716 + 6731) diagram is very interesting, revealing the eight 
regions
to be neatly separated and ordered in pairs along the PN
band. This is in line with the trends found in other PNe (see the case
of M~2-9, Phillips \& Cuesta 1999): the rim appears on the right 
of the diagram, whereas the outer zones of the nebula lie in the
bottom left of the diagram. An interesting result from this
diagram is that the jets and rim, which are high-excitation zones are on the right
side, while both pairs of knots appear to be separated from the former in
terms of degree of excitation.

The bottom panel of Figure~\ref{P-shockphot}, showing
\oiii(4959 + 5007)/\ha\ versus \sii(6716 + 6731)/\ha, separates the zone
of radiatively excited emission lines (the PN zone, adapted from
Phillips \& Cuesta 1999) from the zone mainly excited by shocks (from
the predictions of the plane--parallel and bow-shock models of Hartigan,
Raymond, \& Hartmann 1987).  It is clear that all eight structures in NGC 7009 are
mainly radiatively excited by the PN central star. This is remarkable,
since other PNe show, at least in some of their morphological
components, indications of shock excitation, whereas in NGC 7009 not
even the jet and outer knots approach the shock area in
Figure~\ref{P-shockphot}. On the other hand, the fact that forbidden
lines are radiatively excited in this nebula validates our
calculations of physico-chemical parameters for the different regions.
    
\subsection{Chemical Abundances}

\begin{figure}[ht]
\plotone{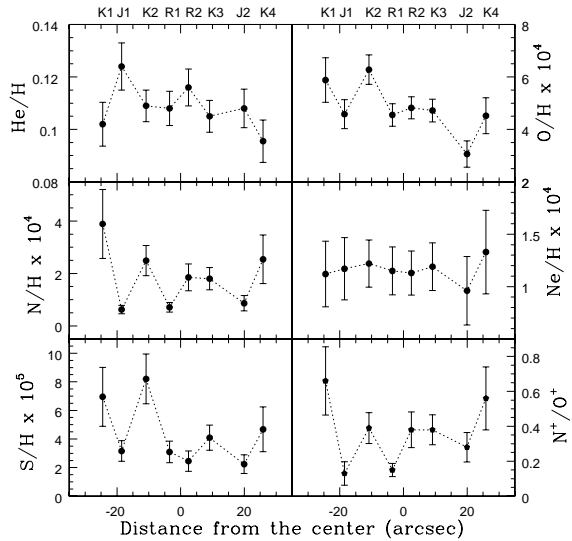}
\caption[ ]{Total abundances profiles for He, O, N, Ne and S. The 
bottom-right panel shows the N$^+$/O$^+$ profile. 
The position of all the  features selected from Figure~1 are marked in the upper 
part of the plot. 
\label{P-Anew}}
\end{figure}

The chemical composition of a PN results
from the mixing of the elements produced by the central star and
dredged up to the surface with those present in the original gas from which the
star was made. The abundances of O, Ne, Ar, and S are thought to
represent the chemical composition of the interstellar medium when the
progenitor star was born, since these elements are neither produced by
the PN progenitor nor very much affected by  nucleosynthesis (see
Iben \& Renzini 1983 and Stasi\'nska 2002 for reviews).  On the
contrary, He, C, and N changes dramatically in the progenitor star 
enriching the stellar surface and, owing to
the dredge-up episodes, enrich the PN envelopes.

Abundances of ionized nebulae are calculated based either on empirical
methods or on model fitting (Stasi\'nska 2002). In the former case,
complete multiwavelength data would give the emission for all ions of
a given element, and the total abundance is the sum of the ionic
ones.  Usually, however, only a few lines of some ions can be measured
and one must correct for unseen ions using ionization correction
factors. 

Here, we have computed the ionic and total abundances (relative to
hydrogen) for the different regions of NGC~7009, and also for the
whole nebula via the ionization
correction factors ({\it icf}) following the scheme in Kingsburgh
\& Barlow (1994), as done, for example, by Corradi et al.~(1997).  The
uncertainties in this method have been discussed by Alexander \& Balick
(1997) and appear to be significant for the analysis of spatially
resolved long-slit observations.

Table~3 lists the ionic and total abundances and the {\it icf}
used. Total abundance profiles along the major axis of the nebula are
shown in Figure~\ref{P-Anew}. Errors in the ionic abundances are
derived by taking into account both the errors in the line ratios and
those in the adopted temperatures. Errors in the total abundances are
obtained by propagating the errors in the ionic abundances as
well as on the {\it icf}. Typical resulting errors (1$\sigma$) in the
total abundances are less than 10$\%$ for He (for which no {\it icf} is
applied), 10--20$\%$ for O, less than 35$\%$ for Ne, and
25--40$\%$ for N and S.

First of all note in Figure~\ref{P-Anew} that the abundances obtained
for the regions \ji\ and \jii\ should be interpreted with great
caution because these are indeed very faint regions in the
nebula. In particular, the \nii\ $\lambda$5755 {\AA} line could not be
measured at the jet positions, and we could not determine
$T_e$\nii. Instead, we have assumed $T_e$\nii\ = $T_e$\oiii\ for \ji\ and \jii,
based on the fact that the $T_e$\nii\ to $T_e$\oiii\ ratio of all the
other regions ranges between 1.01 and 1.25.  Therefore, although the adopted
temperature for \ji\ and \jii\ seems reasonable, the ionic abundances of the
low-ionization ions for these regions are probably more uncertain than
the quoted errors.

Figure~\ref{P-Anew} shows that the abundance of Neon remains constant 
throughout the nebula. Oxygen and Sulfur abundances appear larger in \ki\ 
and \kii\ than in the other regions, but the variations are under the 
2$\sigma$ level.  The constancy of these elements is expected from stellar 
evolution theory and is consistent with previous results for most spatially 
resolved PNe studied so far (cf.~Perinotto \& Corradi 1998). Helium also 
remains practically constant, implying no or little enrichment 
in the nebular ejecta.

Figure~\ref{P-Anew} also shows evidence for an increase of the N abundance 
in the outer knots \ki\ and \kiv, as previously suggested by
Balick et al. (1994). They reported a much higher overabundance, up to a 
factor of 5 for both \kiv\ and \kiii\ (their western ansa and cap), although 
they also warned that their uncertainties in the abundances could be as 
high as a factor of 2.

Our results can also be compared with those of other authors, who
found ``peculiar'' abundances of N and O for certain zones of
NGC~7009. An overabundance of N/H in the outer knots of a factor of 2 or
more with respect to the other regions of the nebula was first derived by
Czyzak \& Aller (1979). Baker (1983), on the other hand, found that 
the N abundance does not vary substantially through the PN, but that O/H is 
marginally (by a factor of $\sim 2$) lower at the position of 
the outer knots. Note, however, that based on much better data and an analysis 
of four PNe containing ansae and including NGC~7009, Balick et al.~(1994)
concluded that {\it only} N/H increases at the ansae, by a factor between 
2 and 5 as compared with the rim. Such behavior was interpreted 
as the result of the recent ejection of N-enriched high-velocity material by 
the PN central star.

A good test to investigate further this point is the inspection of the
profile of the N$^+$/O$^+$ ionic abundance ratio (commonly taken as a
measure of the total N/O abundance that does not depend on {\it icf}
corrections). This is shown in the bottom right panel of Figure~\ref{P-Anew}, 
and shows the same 
trend as the N/H ratio. N/O ratios deduced by Czyzak \& Aller (1979) 
also show an apparent
gradient from the rim to the ansae. Note, however, that Perinotto \&
Corradi (1998) and Hajian et al. (1997) did not find any evidence of peculiar N
or O  abundances in the 17 PNe studied, some of which possess 
ansae.

Finally, our total abundances of He, O, N, Ne, and S for the whole nebula
(integrated over the entire slit, i.e. those labeled NEB in Table~3) are
in a general agreement with those obtained by Baker (1983), Liu et 
al.~(1995), and Hyung \& Aller (1995). More accurate abundances were derived by 
Pottasch (2000) combining IUE and ISO/IRAS line fluxes with optical data from 
the literature, and are also listed in Table 3. These abundances have {\it icf} 
of order of unity and it is therefore relevant to compare them with our values. 
Percentual differences (from 4$\%$ for He to 30-40$\%$ for N and Ne) are similar 
to our estimated errors and indicate a fair level of agreement between the two 
sets.

We find that the sulfur abundance for the whole nebula is lower than the 
corresponding solar value, while the He, O and Ne ones are approximately solar 
(Grevesse \& Anders 1989; Allende-Prieto et al.~2001); the N abundance appears 
similar or marginally higher than the solar value. 

In conclusion, we find a homogeneous elemental abundance across the
nebula, to within 9\%, 17\%, and 35\% for He, O, Ne and S,
respectively.  Nitrogen seems to be enhanced in the outermost knots of NGC~7009 
by a factor $\lesssim 2$, but this evidence is only marginal considering the 
errors and the enormous range in the derived {\it icf}. The uncertainties 
intrinsic to the method are also rather large (Alexander \& Balick 1997), 
but the present data seem to discard variations at the level found by Balick 
et al. (1994).

\section{Discussions}

How the observed physico-chemical characteristics of the
microstructures in NGC 7009 fit into current models for LIS formation
and evolution?  Let us first note that discrete knots such as \ki\ to 
\kiv\ could form during
the evolution of continuous jets: dense knots develop at the tips of
jets as matter accumulates at the  head of the jet, while knots in the main body of the
jet can originate via several forms of instabilities (kink, Vishniac,
Rayleigh--Taylor, and Kelvin--Helmholtz). Sometimes, both jets and knots
are observed, whereas in other instances we observe only  the latter.
From a theoretical point of view, the cooling conditions on the
post-shock gas and the stage of the jet's evolution mainly determine
whether or not both jets and knots are seen (see, for instance,
Soker 1990; Frank, Balick \& Livio 1996 and Garc\'\i a-Segura 1997).

According to hydrodynamical (HD) and magnetohydrodynamical (MHD)
studies (Garc\'\i a-Segura et al. 1999; Garc\'\i a-Segura \& L\'opez
2000; Steffen, L\'opez, \& Lim 2001; Gardiner \& Frank 2001; Blackman et
al.~2001), the interplay between the stellar AGB and
post-AGB winds (for single-star progenitors) or between stellar and
disk winds (if the progenitor is a binary star) can form highly
collimated jets in PNe. Jets that originate in this way are
supersonic, highly collimated, two-sided, and roughly coeval with the
main shell (single star) or younger (binaries). Another characteristic
of some of the models is the linear increase of the jets' expansion
velocity with  distance from the center (Garc\'\i a-Segura et
al. 1999; Steffen et al.~2001). The orientation of jets that
result from these models depends on the alignment between the shell
axis and the disk/rotation (or magnetic) axis (Garc\'\i a-Segura \&
L\'opez 2000; Blackman et al.~2001), as well as on the
presence of precession or wobbling (Cliffe et al. 1995; Garc\'\i
a-Segura 1997; Livio \& Pringle 1997).

If the kinematical results of Reay \& Atherton (1985) are correct (but
see the discussion in Section 3), then, as we have shown in a previous paper
(Gon\c calves et al.~2001),  the morphology and
kinematics of the jet and knots of NCG~7009 can be qualitatively
explained in the light of  HD or MHD single-star
interacting stellar wind model. However, as far as the comparison of our
results with models in terms of physical parameters is concerned, the
situation becomes worse. On the one hand, most of the above models
do not show clear predictions for the densities and temperatures of jets
and knots. On the other hand, those which do show such predictions are
hardly in agreement with the physical parameters derived from the
present observations, as we will show in the following.

The simple process of mass accumulation implies that knots at the 
jets' tips are necessarily denser than the jets themselves. As we showed in 
Section~4 this is not the case in NGC~7009. An additional, very 
intriguing, property of the jets/knots of this PN is that we find no
evidence for shock excitation from the optical lines
(the line ratios of the jets and outer knots appear to be essentially radiatively 
excited, see Section 4.3). 

Further insights can be gained by comparing our results with the
following detailed models.  Steffen et al.~(2001) developed
HD stagnation knot/jet 2-D models which are based on the slow AGB and fast
post-AGB winds interaction with lack of momentum in the polar
direction of the main shell. They show that this interaction gives
rise to ``microstructures'' which first produce a slightly elongated knot 
attached to the rim, then a knot
further elongated and detached to the rim, and finally a polar jet with
a characteristic outward linear increase in velocity. The
stagnation knot/jet model has transient lower density jets with higher
density knots at the tips, but the opposite occurs with the more
permanent jet structures. Therefore, the stagnation knot/jet models
do not explain the outer knots and jets of NGC~7009, whose outer pair
of knots and jets have very similar densities, contrary to the model
predictions.

Recent 3D MHD models by Garc\'\i a-Segura \& L\'opez (2000) can also
be qualitatively compared with our results. These models seem to be
able to reproduce successfully the formation of large- and small-scale
structures in PNe. These authors state that the mass-loss rate during
the post-AGB is a key parameter in distinguishing between the formation
of PNe with highly collimated jets and PNe with knots (ansae-like
structures): $\dot{M} \gtrsim 10^{-7}$~$M_{\odot}$~yr$^{-1}$ leads to the
formation of jets, while lower mass-loss rates ($\dot{M}\sim
10^{-8}$~$M_{\odot}$~yr$^{-1}$) are only able to produce knots, instead 
of jets. Although the low resolution of these models
precludes making quantitative predictions for the physical parameters
of the knots and jets, their results suggest (their Figure~7) that
jets and knots are indeed shock-dominated while our analysis indicates
that the jets and the outer knots of NGC~7009 are mainly radiatively
excited.

An alternative possibility for the formation of polar knots/jets in
elliptical PNe is the model proposed by Frank et al.~(1996). 
They suggest that jets and ansae can be formed during the
``momentum-conserving'' proto-PN stage, i.e., before the beginning of
the fast wind. The key ingredients of this model are the gradual
change from a slow to a fast wind, as well as an appreciable 
equator-to-pole density contrast in the slow wind. The accelerated wind would
flow preferentially parallel to the shock, and the gas would follow the
shocks' prolate boundary toward the polar axis.  The final result is
that the gas flows converge at the polar axis where they collide, form
a new shock that redirects the flow into a jet, and moves outward along
the main axis of the bubble. Jets/ansae formed this way can be
accelerated up to 70\kms\ and have densities $> 10^6$~cm$^{-3}$,
before being ionized by the UV photons of the post-AGB star.

In fact the jet/ansae of NGC~7009 look like a continuity of its
elliptical rim (remembering that rim and jet are very similar in
excitation degree), which is very much in the line of the ``converging
polar flows'' model for the formation of jets and ansae. However, such
flows do not tend to converge into stable ``microstructures'', since the
fast/slow wind interaction could destroy these structures (Frank
et al.~1996; Dwarkadas \& Balick 1998; L\'opez 2002).

Finally, if the nitrogen overabundance of the outer pair of knots is
taken as real (but see the discussion in Section 4.4), the formation of
\ki\ and \kiv\ might be related to recent high-velocity ejections of
enriched material from the central star.  
Such a conclusion requires that the high-velocity
nature of the structures is effectively confirmed by more robust
kinematical data (especially concerning the proper motions).

Another issue to be further investigated is the relation between the
thin jet and the ``fat'' polar bubbles of the nebular shell. The jet
seems to be located along the polar axis of the bubbles, and the outer
knots lie exactly at the edge of the bubbles. It is then conceivable
that the outer knots are the results of the interaction of the jet
with the walls of the bubbles.  However, would such interaction enhance the
low-ionization emission without developing a significant density
contrast, as observed?

The formation of the inner system of knots is also
uncertain. Morphologically, the knots seem to be located in the shell,
and might well share its expansion velocity, as found in other PNe,
such as IC 2553 and NGC~5882 (Corradi et al. 2000). If so, they could
well be the results of in situ instabilities, or survivors of
pre-existing condensations in the AGB wind. Our suggestions for the
formation of the outer and inner pair of knots can readily be tested
with  adequate kinematical modeling of the different structures in
NGC~7009.

\section{Conclusions}

A physico-chemical investigation of the symmetrical pairs of structures 
-- rim, inner and outer knots, and jetlike systems connecting the latter -- in 
NGC~7009 is presented. This study, the most extensive (eight individual regions 
along the major axis), better resolved (1 arcsec spatial resolution 
within the central 1 arcmin) and complete to date (including all 
diagnosis from available optical lines), has 
revealed a clear and coherent pattern of physical, excitation, and chemical 
characteristics through the nebula: 1) the electron temperature, both for 
low- and high- excitation species, is remarkably constant; 2) electron 
densities are similar for both the jetlike structures and the outer knots, 
but they are roughly one-third of those found in the pair of inner knots 
and the rim; 3) no sign of shock excitation is found for any of the 
microstructures or the rim; and 4) a notable chemical uniformity is found 
for the eight regions studied, with the exception of a marginal, moderate 
($\times$ 2) enhancement of nitrogen at the outernmost knots.

The origin of the system of microstructures in NGC~7009 is discussed in the 
light of its physico-chemical parameters and the available theoretical models, 
but none of these appears able to account for both the lack of a strong 
density contrast between the outer knots and jetlike structures, and the 
absence of shocks in them. More than 200 years after its discovery by Herschel, the 
Saturn Nebula is still challenging our comprehension.

\section{Acknowledgments}

We would like to thank the referee, Letizia Stanghellini, for several useful 
comments and suggestions that had help us to improve the paper. We are grateful 
to F. Sabbadin for detecting a misidentification (now corrected) of features in Fig. 1 
of the preprint. The work of DRG, RLMC and AM is partially supported by a 
grant from the Spanish Ministry of Science and 
Technology (AYA 2001-1646).

\clearpage

\pagestyle{empty}
\begin{deluxetable}{llllllllll}
\rotate
\tablenum{1}
\tablewidth{16.5truecm}
\tablecaption{Measured line fluxes normalized to \hb =100.}
\tablehead{
\multicolumn{1}{l}{Line identification}& 
\multicolumn{1}{c}{\ki}& 
\multicolumn{1}{c}{\ji}&
\multicolumn{1}{c}{\kii}&
\multicolumn{1}{c}{\ri}&
\multicolumn{1}{c}{\rii}&
\multicolumn{1}{c}{\kiii}&
\multicolumn{1}{c}{\jii}&
\multicolumn{1}{c}{\kiv}&
\multicolumn{1}{c}{NEB}}
\startdata
{}[O{\sc ii}] 3726.0 + 3728.8	& 		182	    &    	 	63.5	    &    	27.3	    &    	13.8	    &    	6.52	    &    	24.8	    &    	27.0	    &    	138	    &    	21.5\\
H12 3750.15 	 	&	  		-	    &      		1.87	    &      	1.28	    &      	2.12	    &      	1.64	    &      	2.05	    &      	3.53	    &      	1.72	    &      	2.05\\
H11 3770.6             	& 	 		2.14	    &      		3.25	    &      	1.65	    &      	2.31	    &      	2.10	    &      	2.53	    &      	3.95	    &      	2.35	    &      	2.48\\
H10 3797.9           	& 			2.07	    &     		3.37	    &     	2.77	    &     	3.67	    &     	3.05	    &     	3.68	    &     	5.52	    &     	3.49        &     	3.69\\
H9  3835.4	     	& 	 		3.08	    &      		7.90	    &      	4.49	    &      	6.01	    &      	5.40	    &      	6.21	    &      	10.2	    &      	7.35	    &      	6.10\\
{}[Ne{\sc iii}] 3868.7	  &        		74.0	    &       		114	    &       	74.2	    &       	95.5	    &       	94.4	    &       	99.4	    &       	150	    &       	127	    &       	97.1\\
He{\sc i} 3888.7 + H8 3889.1   &	        12.9	    &      		25.9	    &      	15.4	    &      	20.0	    &      	14.6	    &      	19.3	    &      	32.2	    &      	22.6	    &      	18.8\\
{}[Ne{\sc iii}] 3967.5 + H$\epsilon$ 3970.1&   	32.0	    &        		53.5	    &        	36.2	    &        	46.9	    &        	40.5	    &        	46.1	    &        	70.5	    &        	54.2	    &        	44.4\\
He{\sc i} 4026.0	 &	   		1.70	    &       		2.94	    &       	1.90	    &       	2.32	    &       	2.00	    &       	2.35	    &           3.78	    &       	2.66	    &       	2.23\\
{}[S{\sc ii}] 4068.6	 &	   		6.81	    &       		3.20	    &       	1.71	    &       	1.38	    &       	2.17	    &       	2.15	    &       	2.63	    &       	4.62	    &       	1.82\\
{}[S{\sc ii}] 4076.4	 &	   		2.17	    &    	 	1.53	    &    	0.78        &           0.80        &    	1.07	    &    	0.98        &    	0.84	    &    	1.68	    &    	1.02\\
H$\delta$ 4101.8	 &	   		14.9	    &      		30.0	    &      	21.0	    &      	28.0	    &      	22.3	    &      	28.0	    &      	40.5	    &      	27.8	    &      	27.0\\
He{\sc i} 4120.          &     	   		-	    &      		-	    &      	-	    &      	0.49        &      	-	    &      	-	    &      	-	    &      	-	    &      	0.52\\
O{\sc ii} + He{\sc i} 4143.7  &  		-	    &     		-	    &     	0.28        &     	0.41        &     	-	    &     	-	    &     	-	    &     	-	    &     	0.41\\
{}[Fe{\sc v}] 4228.       &         		-	    &      		-	    &      	-	    &      	0.26        &     	-	    &      	0.31        &      	-	    &      	-	    &      	-\\
C{\sc ii} 4267.2	 &	   		-	    &       		-	    &       	0.30        &       	0.64        &       	0.76	    &       	0.68        &       	0.47        &       	-	    &       	0.59\\
H$\gamma$ 4340.5	 &	   		29.2	    &      		56.1	    &      	37.9	    &      	48.1	    &      	40.5	    &      	48.5	    &      	67.2	    &      	48.5	    &      	46.8\\
{}[O{\sc iii}] 4363.2	 &	   		6.57	    &        		9.40	    &        	5.98	    &        	7.35	    &        	8.19	    &        	7.78	    &           13.0	    &        	9.13	    &        	7.72\\
He{\sc i} 4387.9	 &	   		-	    &       		-	    &       	0.65        &      	0.64        &       	-	    &       	0.55        &           -	    &       	-	    &       	0.65\\
He{\sc i} 4471.5	 &	   		3.71	    &       		6.07	    &       	4.20	    &       	5.01	    &       	4.27	    &       	5.11	    &       	7.57	    &       	5.84	    &       	4.85\\
He{\sc ii} 4541.7        &	   		-	    &    	 	-	    &    	0.19	    &    	0.45        &    	-	    &    	-	    &    	-	    &    	-	    &    	0.35\\
N{\sc iii} + O{\sc ii} 4641.0&  		-	    &      		1.50	    &      	-	    &      	3.07	    &      	4.61	    &      	2.68	    &      	1.99	    &      	-	    &      	3.09\\
C{\sc iii} 4647.1	 &	   		-	    &      		1.28	    &      	0.59        &      	1.13	    &      	-	    &      	0.92        &      	1.33	    &      	-	    &      	1.04\\
He{\sc ii} 4685.7        &         		0.99	    &     		2.62	    &     	10.2	    &     	25.6	    &     	23.0	    &     	8.42	    &     	2.16	    &           1.21	    &     	15.6\\
{}[Ar{\sc iv}] + He{\sc i} 4711+13 &     	2.16	    &      		4.62	    &      	3.23	    &      	5.40	    &      	4.89	    &      	2.78	    &      	5.08        &      	1.97	    &      	4.04\\
{}[Ar{\sc iv}] 4740.2    &     		        1.13	    &       		3.63	    &       	2.78	    &       	5.50	    &       	4.49	    &       	2.54	    &       	3.95	    &       	1.48	    &       	3.88\\
H$\beta$ 4861.3	      &            		100	    &      		100	    &      	100	    &      	100	    &      	100	    &      	100	    &      	100	    &      	100	    &      	100\\
{}[O{\sc iii}] 4958.9	&           		428	    &        		473	    &        	435	    &        	405	    &        	428	    &        	424	    &        	458	    &        	454	    &        	421\\
{}[O{\sc iii}] 5006.86  &          		1256	    &       		1339	    &       	1268	    &       	1179	    &       	1243	    &       	1227	    &       	1381	    &       	1329	    &       	1223\\
{}[Ar{\sc iii}] 5191.8  &          		-	    &       		-	    &       	0.09        &       	-	    &       	-	    &       	-	    &       	-	    &       	-	    &       	0.07\\
{}[N{\sc i}] 5200.2    &          		6.06	    &    	 	-	    &    	0.06        &    	-	    &    	-	    &    	-	    &    	-	    &    	2.82        &    	0.11\\
{}[Fe{\sc iii}] 5270.    &          		-	    &      		-	    &      	-	    &      	-	    &      	-	    &      	-	    &      	-	    &      	-	    &      	0.08\\
{}[Fe{\sc iii}] + [Ca{\sc v} 5292.  &		-	    &      		-	    &      	-	    &      	-	    &      	-	    &      	-	    &      	-	    &      	-	    &      	0.09\\
He{\sc ii} 5411.6           &        		-	    &     		-	    &     	0.72        &     	1.48	    &     	1.67	    &     	0.95        &     	-           &     	-	    &     	1.12\\
{}[Cl{\sc iii}] 5517.7        &     		-	    &      		-	    &      	0.71	    &      	0.46        &      	0.46        &      	0.55        &      	0.91        &      	1.03	    &      	0.58\\
{}[Cl{\sc iii}] 5537.9        &     		-	    &       		0.40	    &       	0.85        &       	0.57        &       	0.59	    &       	0.71        &       	0.79        &       	0.96        &       	0.68\\
{}[O{\sc i}] 5577.4          &  		0.51	    &      		-	    &      	0.14	    &      	-	    &      	-	    &      	-	    &      	-	    &      	-	    &      	0.01\\
{}[O{\sc iii}] 5592.4         &    		-	    &        		-	    &        	-	    &        	0.03        &        	0.40        &        	-	    &        	-	    &        	-	    &        	0.05\\
{}[N{\sc ii}] 5754.6	      &    		6.95	    &       		-	    &       	0.72        &       	0.15        &       	0.20	    &       	0.69        &       	-	    &       	4.19	    &       	0.50\\
He{\sc i} 5875.7              &   		20.3        &       		17.0	    &       	17.9	    &       	15.0	    &       	15.3	    &           15.8	    &       	15.8	    &       	16.2	    &       	15.7\\
{}[K{\sc iv}] 6102.          &      		-	    &    	 	-	    &    	-	    &    	0.16        &    	0.18        &    	0.11        &    	-	    &    	-	    &    	0.15\\
He{\sc ii} 6166 + 6170   &      		-	    &      		-	    &      	-	    &      	-	    &      	-	    &      	-	    &      	-	    &      	-	    &      	0.06\\
He{\sc ii} 6234.         &      		-	    &      		-	    &      	-	    &      	-	    &      	-	    &      	-	    &      	-	    &      	-	    &      	0.06\\
{}[O{\sc i}] 6300.3	     &      		31.0	    &     		-	    &     	1.02	    &     	-	    &     	-	    &     	1.69	    &     	-	    &     	14.8	    &     	1.06\\
{}[S{\sc iii}] 6312.1	     &      		4.29	    &      		1.80	    &      	2.84	    &      	1.41	    &      	1.26	    &      	1.97	    &      	1.50	    &      	3.59	    &      	1.86\\
Si{\sc ii} 6347.1           &       		-	    &       		-	    &       	-	    &       	0.07        &       	-	    &       	-	    &       	-	    &       	-	    &       	-\\
{}[O{\sc i}] 6363.8         &       		11.3	    &      		-	    &      	0.38        &      	0.12        &     	0.17        &      	0.56        &      	-	    &      	5.33	    &      	0.36\\
Si{\sc ii} 6371.            &       		-	    &        		-	    &        	-	    &        	0.10        &        	0.12        &        	-	    &        	-	    &        	-	    &        	-\\
He{\sc ii} 6406.5            &       		-	    &       		-	    &       	-	    &       	0.19        &       	0.19        &       	-	    &       	-	    &       	-	    &       	-\\
{}[N{\sc ii}] 6548.0         &      		132	    &       		10.8	    &       	17.9	    &       	3.57	    &       	3.57	    &       	13.9	    &       	7.05	    &       	72.5	    &       	11.2\\
H$\alpha$ 6562.8		&   		394	    &    	 	337	    &    	 408	    &    	312	    &    	341	    &    	280	    &    	217	    &    	 224        &    	313\\
{}[N{\sc ii}] 6583.4	 	&  		397	    &      		19.5	    &      	53.8	    &      	7.93	    &      	6.26	    &      	39.2	    &      	14.7	    &      	217	    &      	30.2\\
He{\sc i} 6678.1	 &	 		8.41	    &      		5.52	    &      	6.16	    &      	4.35	    &      	4.44	    &      	4.00	    &      	3.74	    &      	3.57	    &      	4.46\\
{}[S{\sc ii}] 6716.5	 &	 		41.4	    &     		2.52	    &     	4.82	    &     	0.54        &     	0.43        &     	3.13	    &     	2.59	    &     	26.1	    &     	2.63\\
{}[S{\sc ii}] 6730.8	 &	 		57.4	    &      		3.09	    &      	8.19        &      	0.95        &      	0.78        &      	5.42	    &      	3.05	    &      	32.0	    &      	4.34\\
\hline
\\
$N_e$\sii       & 2000    & 1300    & 4500    & 5500    & 5900    & 5000    & 1100    & 1300    & 4000\\
$N_e$\cliii     & -       & -       & 4700    & 5200    & 5900    & 6000    & 1300    & 1900    & 4500\\
$T_e$\oiii      & 9600    & 10\,400 & 9300    & 10\,000 & 10\,200 & 10\,100 & 11\,600 & 10\,400 & 10\,100\\
$T_e$\nii       & 11\,000 & -       & 9400    & 10\,400 & 12\,800 & 10\,400 & -       & 11\,700 & 10\,300\\
$T_e$\sii       & 7100    & -       & 8300    & -       & -       & -       & -       & 9400    & -\\
\enddata
\end{deluxetable}

\clearpage

\newpage
\pagestyle{empty}
\begin{deluxetable}{lccccccccc}
\rotate
\tablenum{2}
\tablewidth{16.5truecm}
\tablecaption{Estimated percentage errors in line fluxes in Table~1 and 
\hb\ fluxes}
\tablehead{
\multicolumn{1}{l}{Line fluxes}& 
\multicolumn{1}{c}{\ki}& 
\multicolumn{1}{c}{\ji}&
\multicolumn{1}{c}{\kii}&
\multicolumn{1}{c}{\ri}&
\multicolumn{1}{c}{\rii}&
\multicolumn{1}{c}{\kiii}&
\multicolumn{1}{c}{\jii}&
\multicolumn{1}{c}{\kiv}&
\multicolumn{1}{c}{NEB}\\
 & 4.9\tablenotemark{a} & 4.2\tablenotemark{a} & 8.4\tablenotemark{a} &
 3.5\tablenotemark{a} & 1.4\tablenotemark{a} & 9.1\tablenotemark{a} &
 4.9\tablenotemark{a} & 4.9\tablenotemark{a} & 55.3\tablenotemark{a}}
\startdata
(0.01--0.05)I$_{{\rm H}\beta}$ &26 &31 &5  & 6  & 10 & 4.5& 11 & 13.5& 4\\
(0.05--0.15)I$_{{\rm H}\beta}$ &14 &10 &3.5& 4  & 6.5& 3.5& 7.5& 10  & 3.5\\
(0.15--0.30)I$_{{\rm H}\beta}$ &9  &8  &3.5& 3.5& 4.5& 3.5& 6  & 7.5 & 3\\
(0.30--2.0)I$_{{\rm H}\beta}$ &6  &6  &3.5& 3.5& 3.5& 3  & 4.5& 5.5 & 3\\
(2.0--5.0)I$_{{\rm H}\beta}$  &5  &4.5&3  & 3.5& 3.5& 3  & 4  & 5   & 3\\
(5.0--10.0)I$_{{\rm H}\beta}$ &4.5&4  &3  & 3  & 3  & 3  & 4  & 5   & 3\\
$>$ 10 I$_{{\rm H}\beta}$       &4  &4  &3  & 3  & 3  & 3  & 4  & 5   & 3\\
\hline
\\
I$_{{\rm H}\beta}$\tablenotemark{b}  & 1.05(6)  & 1.24(6)   & 30.2(4)  & 30.7(4)   & 15.5(4)   
& 70.2(3)   & 1.60(5)  & 1.28(6)   & 188(3)\\
\enddata
\tablenotetext{a}{Size (in arcsec) of the selected structure along the slit.}
\tablenotetext{b}{In units of $10^{-13}$ erg cm$^{-2}$ s$^{-1}$. Values within 
brackets are the estimated percentage errors in the \hb\ line emission.}
\end{deluxetable}

\clearpage

\newpage
\pagestyle{empty}
\begin{deluxetable}{lllllllllll}
\tabletypesize{\scriptsize}
\rotate
\tablenum{3}
\tablewidth{23.5truecm}
\tablecaption{Ionic/total abundances\tablenotemark{1}}
\tablehead{
\multicolumn{1}{l}{}& 
\multicolumn{1}{l}{\ki}& 
\multicolumn{1}{l}{\ji}&
\multicolumn{1}{l}{\kii}&
\multicolumn{1}{l}{\ri}&
\multicolumn{1}{l}{\rii}&
\multicolumn{1}{l}{\kiii}&
\multicolumn{1}{l}{\jii}&
\multicolumn{1}{l}{\kiv}&
\multicolumn{1}{l}{NEB}&
\multicolumn{1}{l}{NEB - Pottasch}}
\startdata
 He$^+$/H   	&    1.02E-1(06)&  1.23E-1(05)  &    1.04E-1(04)&    9.54E-2(04)&	9.18E-2(04)&
 9.46E-2(04)&    1.07E-1(05) &	 9.55E-2(06)  &	   9.77E-2(04) &  \\
 
 He$^{2+}$/H  	&    - 	  	&  1.2E-3(10)  &    5.02E-3(06)&    1.24E-2(07)&	2.43E-2(06)&	
 1.07E-2(07)&    1.87E-3(08) &	  -           &	   1.30E-2(06)&	  \\
 
 {\bf He/H }    &    {\bf 1.02E-1(09)}&  {\bf 1.24E-1(08)}  &    {\bf 1.09E-1(06)}&    {\bf 1.08E-1(08)}&	
 {\bf 1.16E-1(06)}&	{\bf 1.05E-1(06)}&    {\bf 1.08E-1(08)} &	  {\bf 9.55E-2(09)}&	   
 {\bf 1.11E-1(06)} &	{\bf 1.107E-1}  \\ 
 
 O$^0$/H     	&    4.5E-5(11)&  -  &    2.40E-6(09)&    5.0E-7(13)&	3.1E-7(16)&	
 2.61E-6(08)&    - 	    &	  1.8E-5(13) &	   1.71E-6(08)&	  \\
 
 O$^+$/H    	&    7.5E-5(19)&  2.2E-5(13)  &    2.4E-5(14)&    7.0E-6(15)&	1.3E-6(16)&	
 1.4E-5(14)&    6.06E-6(17)&	  4.2E-5(21)&	   1.2E-5(13)&	 \\
 
 O$^{2+}$/H   	&   5.12E-4(09) &  4.32E-4(08)  &    5.84E-4(06)&    4.12E-4(07)&	4.11E-4(06)&	
 4.25E-4(06)&    2.9E-4(10) &	  4.1E-4(10) &	   4.21E-4(06)&	  \\
 
 {\it icf}(O)   &   1.00  	&   1.00 	&    1.03	&    1.08	&	1.16	   &	
 1.07	   & 	 1.01     &	 1.00        &	   1.08&	  \\
 
 {\bf O/H}      &   {\bf 5.8E-4(15)} &  {\bf 6.4E-4(12)}  &    {\bf 6.28E-4(09)}& {\bf 4.5E-4(11)}&
 {\bf 4.82E-4(09)}& {\bf 4.7E-4(11)}&    {\bf 3.0E-4(15)} &	 {\bf 4.5E-4(17)}  &  {\bf 4.71E-4(09)}
 & {\bf 5.2E-4}	  \\
 
 N$^0$/H     	&   8.7E-6(16) &  -  &    2.2E-7(12)&    - 	     	&	-	   &	
 -	   &	- &	 2.7E-6(19)  &	   2.7E-7(11)&	  \\
 
 N$^+$/H    	&   5.00E-5(08) &  3.09E-6(07)  &    9.66E-6(06)&    1.10E-6(07)&	5.19E-7(08)&	
 5.59E-6(06)&    1.7E-6(10) &	  2.38E-5(09) &	   4.45E-6(06)&	  \\
 
 {\it icf}(N)   &   7.8  	&   20.1 	&    25.7	&    64.7	&	355    &	
 32.1	   &	 50.5    &	  10.6      &	   38.2&	  \\
 
 {\bf N/H}      &   {\bf 3.8E-4(35)} &  {\bf 6.2E-5(26)}  &    {\bf 2.4E-4(24)}&    {\bf 7.0E-5(26)}&	{\bf 1.8E-4(29)}&	
 {\bf 1.8E-4(24)}&    {\bf 8.6E-5(34)}&	 {\bf 2.5E-4(38)}  &   {\bf 1.7E-4(23)}& {\bf 1.3E-4}  \\	
    
 Ne$^{2+}$/H  	&   9.8E-5(14) &  1.1E-4(12)  &    1.1E-4(10)&    1.0E-4(10)&	9.6E-5(10)&	
 1.0E-4(10)&    9.3E-5(17) &	   1.2E-4(15)&	   1.05E-4(09)&	  \\
 
 {\it ic}f(Ne)  &   1.14	&   1.06 	&    1.07	&    1.10	&	1.17	   &	
 1.11	   &    1.03      &	 1.10        &	   1.11&	  \\
 
 {\bf Ne/H}     &   {\bf 1.1E-4(29)} &  {\bf 1.1E-4(26)}  &   {\bf  1.2E-4(20)}&    {\bf 1.1E-4(21)}&	{\bf 1.1E-4(20)}&	
 {\bf 1.1E-4(20)}&    {\bf 9.6E-5(34)} &  {\bf 1.3E-4(30)}  &	   {\bf 1.1E-4(18)}& {\bf 1.9E-4}  \\
 
 S$^+$/H    	&    2.2E-6(13)& 8.99E-8(08)   &    5.11E-7(08)&    4.4E-8(11)&	2.2E-8(13)&	
 2.7E-7(10)&    7.0E-8(10) &	 1.0E-6(14)  &	   2.09E-7(06)&	  \\
 
 S$^{2+}$/H   	&   7.45E-6(21) &  3.2E-6(15)  &    7.4E-6(14)&    2.1E-6(16)&	9.7E-7(19)&	
 3.4E-6(14)&  1.6E-6(19)& 4.8E-6(23)&	   3.3E-6(13)&	  \\
 
 {\it icf}(S)   &    1.43  	&   1.91  	&    2.07	&    2.79	&	4.91	   &	
 2.22	   &    2.58       & 1.57       &	   2.35&	  \\
 
 {\bf S/H}      &   {\bf 1.39E-5(30)} &  {\bf 6.3E-6(23)}  &    {\bf 1.6E-5(23)}&    {\bf 6.1E-6(26)}&	{\bf 4.9E-6(30)}&	
 {\bf 8.1E-6(23)}&  {\bf4.4 E-6(29)}& {\bf 9.3E-6(35)}&	   {\bf 8.3E-6(26)}& {\bf 11.1E-6(26)} \\
\enddata
\tablenotetext{1}{Estimated percentage errors in brackets.}
\end{deluxetable}
\end{document}